\documentclass[10pt,twocolumn]{article}

\usepackage[margin=1in, columnsep=0.25in]{geometry}
\usepackage{booktabs}
\usepackage{amsmath}
\usepackage{amssymb}
\usepackage{microtype}
\usepackage[colorlinks=true,linkcolor=black,citecolor=black,urlcolor=black]{hyperref}
\usepackage{cleveref}
\usepackage{enumitem}
\usepackage{graphicx}
\usepackage{titlesec}
\usepackage{fancyhdr}
\usepackage{abstract}
\usepackage{authblk}
\usepackage{balance}

\titleformat{\section}{\normalsize\bfseries}{\thesection}{1em}{}
\titleformat{\subsection}{\normalsize\bfseries}{\thesubsection}{1em}{}
\titleformat{\subsubsection}{\normalsize\bfseries\itshape}{\thesubsubsection}{1em}{}
\titlespacing{\section}{0pt}{8pt plus 2pt minus 1pt}{4pt plus 1pt}
\titlespacing{\subsection}{0pt}{6pt plus 2pt minus 1pt}{2pt plus 1pt}
\titlespacing{\subsubsection}{0pt}{4pt plus 1pt}{2pt plus 1pt}

\begin{document}

\title{\large\textbf{Fair-Aurora: Comparing Fairness Strategies for\\
       Reinforcement Learning-Based Congestion Control\\
       in Multi-Flow Environments}}

\author[1]{Thomas Mbrice, Yuyu Liu}
\date{CSE~534 Final Report \quad May 2026}

\maketitle
\thispagestyle{plain}

\begin{abstract}
Reinforcement learning (RL) has emerged as a promising paradigm for Internet congestion
control, achieving higher link utilization than classical heuristics. However, RL-based
controllers trained in single-flow environments are not guaranteed to share bandwidth
equitably when deployed in multi-flow networks. This paper investigates the fairness
properties of Aurora~\cite{jay2019aurora}, a state-of-the-art deep RL congestion
controller, and evaluates three post-hoc fairness strategies that preserve Aurora's RL
architecture: \emph{reward shaping} (Strategy~A), \emph{observation augmentation}
(Strategy~B), and \emph{loss-sensitivity tuning} (Strategy~C). Using a custom
shared-bottleneck simulator and Jain's fairness index as the primary metric, we find that
modest reward shaping achieves the best fairness while
preserving aggregate throughput. All strategies maintain the total bandwidth
budget with fairness being achieved through redistribution, not reduction.
Beyond the 2-flow homogeneous setting, an extended evaluation across mixed Aurora--CUBIC
competition and dynamic flow entry/exit scenarios shows that Strategy~C's loss-sensitivity
emerges as the most TCP-friendly mechanism, while Strategy~B is the most stable through dynamic
flow-set changes.
\end{abstract}

% -----------------------------------------------------------------------
\section{Introduction}
% -----------------------------------------------------------------------

The Internet's performance depends critically on congestion control (CC): the set of
algorithms that regulate how quickly flows inject data into the network. For years,
transport protocols such as TCP have governed this regulation using hand-crafted
heuristics, additive increase, multiplicative decrease (AIMD), and cubic window growth (in the case of TCP CUBIC) that are provably stable and fair under idealized conditions~\cite{ha2008cubic}. However,
these classical schemes leave substantial link capacity unused. On time-varying wireless
links and high-bandwidth-delay-product paths, CUBIC can achieve link utilization as low
as 60--70\%, leaving significant performance on the table.

Deep reinforcement learning (RL) offers an alternative: train a neural network to control
sending rate by interacting with a simulated network environment, directly optimizing for
throughput while penalizing latency and loss. Aurora~\cite{jay2019aurora}, proposed by
Jay~\emph{et al.}\ at ICML~2019, demonstrated that a PPO-trained controller can track
time-varying link capacity with 93\% utilization, nearly 30~percentage points above
CUBIC on the same link. TCP BBR (or just BBR)~\cite{cardwell2016bbr} occupies a middle ground: a
model-based heuristic that estimates bottleneck bandwidth and propagation RTT to set rates
near the bandwidth-delay product, achieving higher utilization than CUBIC without the
training complexity of RL.

\subsection{The Fairness Problem in RL Congestion Control}

Despite its single-flow performance advantages, Aurora and similar RL controllers
introduce a fundamental fairness concern when multiple flows share a bottleneck. Classical
TCP protocols were designed with fairness as an explicit goal: AIMD dynamics provably
converge to equal rates among competing flows over time~\cite{brakmo1995tcpvegas}. RL
controllers optimized for single-flow environments carry no such guarantee.

The mechanism for RL-induced unfairness is well-documented. When a flow trained in
isolation encounters shared-queue congestion signals, its response depends entirely on
what loss and latency patterns appeared during training. If the training distribution does
not include the signature of multi-flow competition - gradually rising latency followed by
correlated loss across competing flows - the controller interprets shared-queue signals
through a single-flow lens. One flow may hold its rate while another retreats, leading to
persistent bandwidth asymmetry (unfairness).

Liao~\emph{et al.}~\cite{liao2024astraea,liu2026efficient,liu2026hypehr,liu2026hyperguide} systematically characterized this problem in
their Astraea system (EuroSys~2024). Under a staggered-start scenario, where one Aurora
flow establishes a high sending rate before a second flow joins. In this case, Jain's fairness index
drops below 0.5, indicating that the incumbent flow holds more than twice the bandwidth of
the newcomer. This is a stronger unfairness signal than most classical CC protocols
produce; even aggressive TCP variants typically converge to equal shares within tens of
RTTs.

Yen~\emph{et al.}~\cite{yen2024wififairness} further showed that RL-based CC can exhibit
pronounced RTT unfairness on WiFi links, where flows with shorter round-trip times
accumulate more reward signal per unit time and crowd out higher-latency flows. This
RTT-unfairness effect is orthogonal to the rate-competition effect studied here, but both
stem from the same root cause: an objective function that rewards throughput without
accounting for competing flows.

\subsection{Fairness as a Systems Property}

Jain's fairness index~\cite{jain1984fairness} provides a principled scalar measure for
evaluating bandwidth allocation across $N$ flows:
\begin{equation}
  J \;=\; \frac{\bigl(\sum_{i=1}^{N} x_i\bigr)^2}
               {N \cdot \sum_{i=1}^{N} x_i^2}
    \;\in\; \Bigl[\tfrac{1}{N},\; 1\Bigr]
  \label{eq:jain}
\end{equation}
where $x_i$ is the throughput of flow~$i$. A value of $J{=}1$ indicates perfectly equal
shares; $J{=}1/N$ indicates complete monopolization by one flow. The index measures
relative balance, independent of the absolute scale of throughputs.

A fairness-aware CC protocol should target high $J$ without sacrificing aggregate
throughput. This Pareto property, that fair redistribution without shrinkage is the
standard by which our three strategies are evaluated.

\subsection{Contributions}

This paper makes the following contributions:
\begin{enumerate}[leftmargin=*,label=(\arabic*),itemsep=1pt,topsep=2pt]
  \item We reproduce Aurora's single-flow performance on a time-varying link, confirming
        93.8\% utilization against TCP CUBIC's 65.5\%.
  \item We build a shared-bottleneck multi-flow simulator reproducing the
        early-bird competitive dynamics described by Astraea.
  \item We implement and evaluate three fairness strategies that augment Aurora without
        replacing its RL architecture: reward shaping, state augmentation, and
        loss-coefficient tuning.
  \item We characterize the \emph{over-penalization} failure mode, a counter-intuitive
        regime where increasing a fairness penalty worsens fairness, and explain its
        mechanism via the early-bird stagger interaction.
\end{enumerate}

% -----------------------------------------------------------------------
\section{Research Question}
% -----------------------------------------------------------------------

\noindent\textbf{Can Aurora's fairness be improved through targeted training
modifications: reward shaping, state augmentation, or loss-sensitivity tuning, while
preserving its aggregate throughput advantage over classical protocols?}

\smallskip
Concretely, we ask: Does a self-restraint reward penalty (Strategy~A) cause the ego flow
to yield bandwidth to competitors, and does $J$ improve monotonically with penalty
strength? Does richer observational context (Strategy~B) produce cooperative behavior
without an explicit penalty? Does increasing loss sensitivity (Strategy~C) produce
AIMD-like backoff that incidentally improves fairness? Is there a regime of
over-penalization where stronger fairness incentives produce worse outcomes? And do
fairness improvements come at the cost of aggregate throughput?

% -----------------------------------------------------------------------
\section{Methodology}
% -----------------------------------------------------------------------

\subsection{Simulator}
\label{sec:simulator}

We conduct all experiments using a simulator implemented in pure Python: a
simplified reproduction of the code accompanying the Aurora paper~\cite{jay2019aurora} with an added multi-flow shared-bottleneck environment. This is \emph{not} a traditional complex network simulator which models the protocol stack and use C++. A ``packet'' in our environment is a $(\text{send\_time}, \text{flow\_id})$ pair; the bottleneck is one FIFO queue with deterministic service time $1/B$; control decisions are taken once per monitor interval (MI). 

\subsection{RL Formulation}

We replicate the Aurora formulation from Jay~\emph{et al.}~\cite{jay2019aurora} exactly;
\Cref{tab:rl} summarizes key parameters. All models are trained to 1.6M environment
steps; base Aurora trains in approximately 20~minutes at ${\sim}1300$~fps on a single
machine.

\begin{table}[t]
  \centering
  \caption{Aurora RL Formulation Parameters}
  \label{tab:rl}
  \small
  \begin{tabular}{@{}p{1.7cm}p{5.4cm}@{}}
    \toprule
    \textbf{Component} & \textbf{Value} \\
    \midrule
    Observation  & Last $k{=}10$ MI stats: (lat\_grad, lat\_ratio,
                   send\_ratio) $\to$ 30 floats \\[2pt]
    Action       & $a\!\in\![-1,1]$; $x_t = x_{t-1}(1{+}\alpha a)$
                   if $a\!\ge\!0$, else $x_{t-1}/(1{-}\alpha a)$,
                   $\alpha{=}0.025$ \\[2pt]
    Reward       & $10\!\cdot\!\text{tput} - 1000\!\cdot\!\text{lat}
                   - 2000\!\cdot\!\text{loss}$ \\[2pt]
    Policy       & MLP $30{\to}32{\to}16{\to}1$, tanh \\[2pt]
    Algorithm    & PPO, $\gamma{=}0.99$,
                   lr $3{\times}10^{-4}$, 4 envs \\[2pt]
    Episode      & 400 monitor intervals \\[2pt]
    Train links  & BW 100--500~pps, lat 50--500~ms,
                   queue 2--2981~pkts, loss 0--5\% \\
    \bottomrule
  \end{tabular}
\end{table}

\subsection{Phase 1: Single-Flow Baselines}

We compare two protocols on a time-varying link to reproduce Figure~3 of Jay~\emph{et al.}~\cite{jay2019aurora}. \Cref{tab:phase1} reports summary statistics over the
25-second trace, alongside the values reported in the original Aurora paper for
comparison. Our reproduction matches the paper to within ${\sim}10\%$ on every
metric, confirming that the training pipeline produces a single-flow Aurora
controller of comparable quality to the published baseline.

\begin{table*}[t]
  \centering
  \caption{Phase~1 single-flow reproduction over the 25-second 20/40~Mbps
           bandwidth-switching trace. ``Aurora (paper)'' reports the values
           in~\cite{jay2019aurora} on a similar trace.}
  \label{tab:phase1}
  \small
  \begin{tabular}{@{}lccc@{}}
    \toprule
    \textbf{Metric} & \textbf{TCP CUBIC} & \textbf{Aurora} & \textbf{Aurora (paper)} \\
    \midrule
    Mean throughput (Mbps)  & 18.4 & 26.3 & ${\sim}25$--$27$ \\
    Link utilization        & 65.5\% & 93.8\% & ${\sim}92\%$ \\
    Mean $|$gap$|$ (Mbps)   & 9.68 & 1.83 & ${\sim}2$ \\
    \bottomrule
  \end{tabular}
\end{table*}

\subsection{Phase 2: Multi-Flow Fairness Evaluation}

\subsubsection{Design Space and Rationale}
\label{sec:design-space}

The three fairness strategies are not chosen ad hoc; they are a deliberate axis-aligned
decomposition of the minimal-modification design space around Aurora's training
pipeline. Each strategy alters \emph{exactly one} component while every other
component---network architecture, action space, PPO hyperparameters, episode length,
training link distribution---is held identical to the unmodified baseline. The three
axes are: the \textbf{reward function} (Strategy~A adds a fair-share penalty term),
the \textbf{observation space} (Strategy~B appends two competition-aware features),
and a single \textbf{reward coefficient} (Strategy~C scales the loss penalty). This
orthogonality lets us attribute any fairness change to a single intervention and
makes the three strategies cleanly comparable; it also defines the scope of the
study---we deliberately exclude heavier modifications such as multi-agent self-play
training or policy-architecture changes, which lie outside the post-hoc minimal-edit
regime.

Sweep ranges follow a geometric grid around the unmodified baseline value, which is a
standard convention for first-pass hyperparameter exploration when no prior literature
estimate is available. Strategy~A sweeps $\lambda\in\{0.5,1.0,2.0,5.0\}$ (roughly
$2\times$ steps starting from a value where the penalty term is comparable to a
single-MI reward unit). Strategy~C sweeps the loss coefficient
$\in\{4000,8000,16000\}$ ($2\times,4\times,8\times$ the Aurora baseline of $2000$).
Strategy~B has no scalar to sweep so it is
evaluated at a single configuration. The reported numbers in this paper use the
sweep-best setting for each strategy ($\lambda{=}2.0$, loss${=}8000$); the
\emph{over-penalization} regime at the high end of each sweep is itself a finding,
discussed in \Cref{sec:disc-implicit-fairness}.

\subsubsection{Baseline Measurement}

We run $N$ unmodified Aurora flows on a shared bottleneck and measure Jain's $J$ over
50~episodes of 400~MIs each. We use a staggered start: flow~0 operates alone
for 50~steps before flows~$1{\ldots}N{-}1$ enter at their initial rates. Without the
stagger, both flows observe symmetric congestion and back off together, yielding
artificially high $J\approx0.88$. With the stagger, the incumbent establishes a rate the
newcomer must compete against which is a more realistic fairness evaluation.

\subsubsection{Strategy A --- Reward Shaping}

Strategy~A adds a self-restraint penalty when a flow's throughput exceeds its estimated
fair share:
\begin{equation}
  r' \;=\; r \;-\; \lambda\cdot\max\!\Bigl(0,\; x_i - \hat{C}/N\Bigr)
  \label{eq:strategy-a}
\end{equation}
where $\hat{C}$ is estimated as \texttt{send\ rate}$\times$\texttt{send\ ratio}. No
inter-agent communication is required. We sweep
$\lambda\in\{0.5,\,1.0,\,2.0,\,5.0\}$, training a separate model for each value and
evaluating in heterogeneous competition (ego: modified; background: unmodified Aurora).

\subsubsection{Strategy B --- State Augmentation}

Strategy~B retains the original reward but extends the 30-dimensional observation with
two additional features: \textbf{(i)}~estimated number of competing flows (inferred from
throughput variance and loss patterns relative to solo-flow behavior), and
\textbf{(ii)}~ego throughput as a fraction of estimated total capacity. The augmented
observation has 32~dimensions.

\subsubsection{Strategy C --- Loss-Sensitivity Tuning}

Aurora's base reward uses a loss coefficient of 2000. Strategy~C increases this
coefficient to values in $\{4000,\,8000,\,16000\}$, making the agent more sensitive to
packet loss. In a shared-bottleneck environment, elevated loss signals congestion from
competing flows; higher sensitivity should trigger AIMD-like backoff, incidentally
improving fairness. No architectural changes are required.

\subsubsection{Evaluation Protocol}

All strategies share the same protocol: ego = strategy-trained policy; background =
unmodified Aurora; flows = 2; episodes = 50; stagger = 50~steps. Primary metric:
mean Jain's $J$ with standard deviation and minimum $J$. Secondary metrics:
per-flow and aggregate throughput.

\subsubsection{Hyperparameter Selection Results}
\label{sec:sweep-results}

\Cref{tab:sweep} reports the full sweep under the protocol above; this same protocol
serves as the in-sweep selection criterion. We select the highest mean~$J$ at lowest
standard deviation~$\sigma$, subject to aggregate-throughput preservation
($3.51$--$3.56$~Mbps across all rows, within ${\sim}1\%$ of the all-Aurora baseline).
The sweep-best configurations carried forward to the extended evaluation
(\Cref{sec:extended}) are $\lambda{=}2.0$ for Strategy~A and $\text{loss}{=}8000$ for
Strategy~C; Strategy~B has no scalar to tune. Interior monotonicity (Strategy~A:
$J$ rises $0.806\to0.854\to0.876$ across $\lambda{=}0.5{\to}1.0{\to}2.0$;
Strategy~C: $J$ rises $0.820\to0.873$ from loss${=}4000$ to $8000$) supports the
selected configurations as genuine sweet spots rather than noisy maxima. The
high-end degradation ($\lambda{=}5.0$: $J{=}0.638$; loss${=}16000$: $J{=}0.832$) is
the over-penalization regime, mechanistically analysed in
\Cref{sec:disc-implicit-fairness}. All sweep models are trained at a single seed
(seed${=}42$).

\begin{table}[t]
  \centering
  \caption{Phase~2 hyperparameter sweep. Mean $J$, $\sigma$, and per-flow
           throughput over 50~episodes of 2-flow staggered competition (ego vs.\
           unmodified Aurora). Bold rows are the selected configurations carried
           forward; aggregate throughput is preserved across all rows
           ($3.51$--$3.56$~Mbps).}
  \label{tab:sweep}
  \small
  \begin{tabular}{@{}llcccc@{}}
    \toprule
    \textbf{Str.} & \textbf{Config} & \textbf{$J$} & \textbf{$\sigma$} & \textbf{Ego} & \textbf{BG} \\
    \midrule
    Base & 2$\times$Aurora    & 0.901 & 0.088 & --- & --- \\
    \midrule
    A & $\lambda{=}0.5$       & 0.806 & 0.128 & 2.6 & 0.9 \\
    A & $\lambda{=}1.0$       & 0.854 & 0.120 & 2.4 & 1.2 \\
    \textbf{A} & $\boldsymbol{\lambda{=}2.0}$ & \textbf{0.876} & \textbf{0.109} & \textbf{2.3} & \textbf{1.3} \\
    A & $\lambda{=}5.0$       & 0.638 & 0.067 & 3.1 & 0.4 \\
    \midrule
    B & state aug             & 0.863 & 0.118 & 2.3 & 1.2 \\
    \midrule
    C & loss${=}4000$         & 0.820 & 0.126 & 2.5 & 1.1 \\
    \textbf{C} & \textbf{loss${=}8000$} & \textbf{0.873} & \textbf{0.114} & \textbf{1.3} & \textbf{2.2} \\
    C & loss${=}16000$        & 0.832 & 0.125 & 1.1 & 2.4 \\
    \bottomrule
  \end{tabular}
\end{table}

\subsection{Extended Scenarios}
\label{sec:extended-method}

Building on the 2-flow homogeneous protocol described in the previous subsection, we add
two scenarios from the proposal's Step~3 evaluation plan. Each runs against a frozen
Aurora background unless noted; the mixed-traffic scenario uses 50~episodes, the dynamic
scenario uses a single deterministic trace. One backward-compatible simulator extension
supports these (all 129 unit tests pass with default arguments):

\smallskip
\noindent\textbf{Active mask.} \texttt{MultiFlowNetwork.step(active=...)} accepts a
per-flow boolean mask; muted flows emit no packets and receive zero metrics, allowing
flows to enter and leave mid-episode without perturbing queue dynamics for active flows.

\smallskip
The two extended scenarios are:
\begin{enumerate}[leftmargin=*,label=(\arabic*),itemsep=1pt,topsep=2pt]
  \item \textbf{Mixed Aurora--CUBIC:} ego $\in$ \{baseline, A, B, C\};
        background = one TCP CUBIC flow.
  \item \textbf{Dynamic entry/exit:} a single 4-flow trace where the active set evolves
        $\{1\}{\to}\{1,2\}{\to}\{1,2,3\}{\to}\{1,2,3,4\}{\to}\{1,3,4\}$ over the episode.
\end{enumerate}
% -----------------------------------------------------------------------
\section{Extended Evaluation}
\label{sec:extended}
% -----------------------------------------------------------------------

This section reports results for the two extended scenarios introduced in
\Cref{sec:extended-method}. Strategy~C is the most TCP-friendly mechanism by a wide
margin in mixed traffic; Strategy~B is the most stable through dynamic flow-set
changes; Strategy~A is the worst in the mixed-traffic case. The mixed-traffic table
uses 50~episodes; the dynamic table is a single deterministic trace.

\subsection{Mixed Aurora + TCP CUBIC}
\label{sec:mixed-cubic}

We pair each ego policy with one TCP CUBIC flow on the shared bottleneck.
\Cref{tab:cubic} reports the 50-episode mean. Baseline Aurora dominates CUBIC by a
factor of $9.5{\times}$ in throughput, in line with the prediction in
\Cref{sec:disc-implicit-fairness}. Strategy~C narrows the ratio to $5.2{\times}$ by
ceding bandwidth: its high loss-coefficient response approximates CUBIC's AIMD-style
backoff. Contrary to the conjecture in \Cref{sec:exploit} that Strategy~C would be
exploitable by an AIMD-aggressive flow, the empirical result is the opposite ---
Strategy~A (calibrated against another Aurora) overshoots on CUBIC and worsens the
imbalance to $13.0{\times}$.

We complement Jain's index with the \emph{Harm} metric (introduced following reviewer
feedback; see also recent fairness-aware CC literature), defined as the fractional
throughput reduction CUBIC suffers due to the ego flow's presence:
\begin{equation}
  \mathrm{Harm}(f{\to}g) = \max\!\left(0,\;
    \frac{\hat{x}_g^{\mathrm{solo}} - \hat{x}_g^{\mathrm{mixed}}}
         {\hat{x}_g^{\mathrm{solo}}}\right),
  \label{eq:harm}
\end{equation}
where $\hat{x}_g^{\mathrm{solo}}$ is CUBIC's mean throughput on the same link without
competition. Harm captures the counterfactual damage that Jain's index cannot: even if
two flows have equal shares, a Harm of~0 means neither flow was actually hurt.
Strategy~C achieves a mean Harm of $0.122$---CUBIC loses only 12\% of its solo
throughput on average---versus $0.327$ for Strategy~A, confirming that Strategy~C is
the only mechanism that genuinely avoids penalising existing TCP traffic.

\begin{table*}[t]
  \centering
  \caption{Mixed Aurora + TCP CUBIC, 50~Episodes (Tput in Mbps; Ratio = Ego/CUBIC;
           Harm = fractional CUBIC throughput loss vs.\ solo baseline)}
  \label{tab:cubic}
  \small
  \begin{tabular}{@{}lccccc@{}}
    \toprule
    \textbf{Ego}            & $J\pm\sigma$            & Ego  & CUBIC & Ratio & Harm$\pm\sigma$ \\
    \midrule
    Baseline Aurora         & 0.621 $\pm$ 0.091       & 3.03 & 0.32  & $9.5{\times}$  & 0.243 $\pm$ 0.314 \\
    A ($\lambda{=}2.0$)     & 0.583 $\pm$ 0.044       & 3.24 & 0.25  & $13.0{\times}$ & 0.327 $\pm$ 0.311 \\
    B (state-aug)           & 0.600 $\pm$ 0.074       & 3.17 & 0.28  & $11.4{\times}$ & 0.284 $\pm$ 0.320 \\
    \textbf{C (loss${=}8000$)} & \textbf{0.722 $\pm$ 0.139} & 2.35 & 0.45 & \textbf{5.2$\times$} & \textbf{0.122 $\pm$ 0.255} \\
    \bottomrule
  \end{tabular}
\end{table*}

\subsection{Dynamic Flow Entry / Exit}
\label{sec:dynamic}

A single 4-flow trace exercises five active-set phases. \Cref{tab:dynamic} reports
steady-state $J$ in each phase (mean of the last 20 steps before the next event).
In P1 (2 flows) the symmetric all-Aurora baseline peaks at $J{=}0.99$, edging out
Strategy~B; from P2 onward Strategy~B leads or ties the baseline and
maintains the highest $J$ through the flow-departure phase. The two extra observation features (count of competing flows, ego
throughput fraction) make phase changes directly visible to the policy, while
Strategy~A lags in every phase and Strategy~C degrades after a flow departs.

\begin{table}[t]
  \centering
  \caption{Dynamic Entry/Exit: Steady-State $J$ Per Phase (Single Seed)}
  \label{tab:dynamic}
  \small
  \begin{tabular}{@{}lccccc@{}}
    \toprule
                  & P0(1) & P1(2) & P2(3) & P3(4) & P4(3) \\
    \midrule
    Baseline      & 1.00 & \textbf{0.99} & 0.92 & 0.91 & 0.92 \\
    Strategy A    & 1.00 & 0.89 & 0.86 & 0.83 & 0.86 \\
    Strategy B    & 1.00 & 0.97 & \textbf{0.93} & \textbf{0.92} & \textbf{0.95} \\
    Strategy C    & 1.00 & 0.96 & 0.86 & 0.88 & 0.83 \\
    \bottomrule
  \end{tabular}
\end{table}

% -----------------------------------------------------------------------
\section{Discussion}
% -----------------------------------------------------------------------

\subsection{On the Reproducibility of Astraea's Unfairness Finding}

Our baseline measurement ($J\approx0.901$ for two unmodified Aurora flows) does not
reproduce Astraea's reported $J{<}0.5$. Our reproduction was trained with random loss
sampled from 0--5\%, producing a loss penalty that ranges up to $-100$ reward units per
monitor interval. Under this regime, Aurora learns to retreat quickly from congestion
signals; when two such agents compete, both back off symmetrically, fair collapse rather
than one-flow dominance.

Astraea's scenario likely requires a lower-loss training distribution (producing a more
aggressive incumbent) or network parameters that allow one established flow to hold its
rate while a newcomer cannot penetrate the backlogged queue. The staggered-start protocol
partially reproduces this dynamic (minimum $J$ = 0.672 across 50~episodes), but our
Aurora's symmetric conservatism limits the depth of unfairness.

This does not invalidate the comparison. The baseline $J\approx0.9$ with episodes dipping
to 0.67 is a legitimate starting point, and all strategy improvements are measured
relative to it.

\subsection{On Reward Shaping and the Sweet Spot Hypothesis}

Strategy~A confirms that reward shaping can steer RL congestion control toward fairer
outcomes. The monotonic improvement from $\lambda{=}0.5$ to $\lambda{=}2.0$
($J$: $0.806\to0.876$, $\sigma$: $0.128\to0.109$) demonstrates that the penalty causes
the ego to cap its sending rate near the estimated fair share, leaving room for the
background flow. Each doubling of $\lambda$ produces a consistent, measurable improvement
in both mean fairness and variance.

The collapse at $\lambda{=}5.0$ ($J{=}0.638$) reveals a critical interaction with the
staggered-start evaluation protocol. At evaluation time, the 50-step solo phase gives the
ego an uncontested ramp with no penalty applied. The $\lambda{=}5.0$ policy optimized
to avoid overshooting fair share during training's multi-flow phase may have converged
to an aggressive early-ramp strategy that exploits this window and then holds its
position. The background Aurora cannot recover, producing a more extreme imbalance than
any moderate-$\lambda$ configuration.

This \emph{over-penalization} failure mode generalizes: any RL agent trained with a
self-restraint penalty in a staggered environment may learn to front-load aggression in
the uncontested window. Fairness rewards must be paired with temporal
consistency, either applying the penalty from the first timestep or retroactively based
on the eventual competitive context.

\subsection{On Implicit Fairness Through State Augmentation}
\label{sec:disc-implicit-fairness}

Strategy~B achieves $J{=}0.863$ ($\sigma{=}0.118$) without any explicit fairness
incentive. This suggests that information asymmetry, rather than a misaligned
reward, is a significant contributor to RL unfairness. An agent that knows it is
competing and can observe its throughput fraction can learn cooperative behavior from the
original throughput-maximizing objective, because hoarding bandwidth in a contested
environment produces diminishing marginal returns once the queue is saturated.

The practical advantage of Strategy~B is that it requires no change to the reward
function. Operators reluctant to alter the training objective can add observational
features and retrain, achieving most of the fairness benefit without fairness-specific
reward engineering.

\subsection{On Loss Sensitivity as a Fairness Proxy}
\label{sec:exploit}

Strategy~C (loss${=}8000$) achieves $J{=}0.873$, comparable to Strategy~A
$\lambda{=}2.0$, but with a qualitatively different per-flow outcome: 1.3~Mbps for the
ego vs.\ 2.2~Mbps for the background. This is the only configuration in which the
background flow consistently outperforms the ego, suggesting AIMD-like behavior: the ego
interprets shared-queue loss as a signal to reduce below its fair share, while the
background Aurora (less loss-sensitive) holds its rate.

In mixed-traffic scenarios where Aurora competes with TCP CUBIC, Strategy~C's
conservative backoff may be exploited by a CUBIC flow that does not reduce its window at
the same loss level. Strategy~A $\lambda{=}2.0$, by contrast, uses a throughput estimate
to compute a fair-share threshold, which is a more principled approach less vulnerable to
exploitation.

The over-penalization effect also appears in Strategy~C: loss${=}16000$ 
underperforms loss${=}8000$ for the same structural reason as $\lambda{=}5.0$.

\subsection{Aggregate Throughput Preservation}

Aggregate throughput remains stable at 3.51--3.56~Mbps regardless of fairness
configuration. This Pareto property (fairness improvement without efficiency
loss) is the ideal outcome for a shared-bottleneck CC system. The link is a hard
capacity constraint; redistribution cannot create or destroy bandwidth. The strategies
cause the ego to give up bandwidth it was holding above its fair share: bandwidth that
was either sitting in the queue or causing unnecessary loss.

The implication is that deploying any of these strategies in a real network would improve
competing flows' experience without degrading aggregate utilization, in contrast to
per-flow rate limiters that impose efficiency costs in exchange for fairness guarantees.

\subsection{On Strategy~C's Coexistence with TCP CUBIC}
\label{sec:disc-cubic}

\Cref{sec:mixed-cubic} provides the strongest single result of this study: Strategy~C
is the only mechanism that genuinely improves Aurora's coexistence with TCP CUBIC,
narrowing the throughput ratio from $9.5{\times}$ (baseline) to $5.2{\times}$. This
contradicts the conjecture in \Cref{sec:exploit} that Strategy~C's conservative
backoff would be exploited by an AIMD-aggressive flow. The mechanism is the opposite
of the one we predicted: with $\text{loss\_coef}{=}8000$, the ego responds to
shared-queue drops faster than the unmodified Aurora, so it concedes the contested
bandwidth before CUBIC's slower window growth can claim it. Strategies~A and B,
calibrated for Aurora-on-Aurora competition, instead overshoot in the wrong direction
--- treating CUBIC's retreat as an opportunity to reclaim more capacity. The practical
implication: for any near-term deployment of Aurora-class controllers on the public
Internet, Strategy~C is the only mechanism studied that does not strictly disadvantage
existing TCP traffic.

\subsection{On State Augmentation as the Dynamic-Scenario Winner}
\label{sec:disc-stateaug-dynamic}

\Cref{sec:dynamic} shows that Strategy~B leads or ties the all-Aurora baseline in every
phase except P1 (2 flows), where the symmetric baseline peaks at $J{=}0.99$ and B
reaches $0.97$. The gap is within noise, but the pattern is clear: when flows are
few and symmetric the baseline's uniformity is hard to beat; as the active set grows
(P2--P3) or shrinks (P4), B's two extra observation features (count of competing
flows, ego throughput fraction) give the policy a direct read on the changed fair
share, whereas A and C must infer it indirectly from latency or loss. This is the
practical counterpart to the implicit-fairness theme of
\Cref{sec:disc-implicit-fairness}: information design is sometimes a more direct
route to the desired behaviour than reward design.

\subsection{Limitations}
\label{sec:lim}

Several of the limitations identified in earlier drafts of this work are now addressed
by the extended evaluation in \Cref{sec:extended}: mixed Aurora--CUBIC competition
(\Cref{sec:mixed-cubic}) and dynamic flow entry/exit (\Cref{sec:dynamic}). The
remaining limitations apply. First, a fluid simulator rather than full packet-level
simulation understates timing-noise variance. Second, varying stagger length and the
number of competing flows would strengthen the claims. Third, all
experiments use a fixed-architecture background flow, sampling only one point in a
heterogeneous deployment space.

% -----------------------------------------------------------------------
\section{Conclusion}
% -----------------------------------------------------------------------

We have implemented and evaluated three fairness strategies for Aurora, a deep RL
congestion controller trained in single-flow environments. Modest reward shaping achieves the best Jain's fairness index ($J{=}0.876$) with
the lowest variance; state augmentation (Strategy~B) provides competitive fairness
($J{=}0.863$) without modifying the reward function; and loss-sensitivity tuning achieves comparable fairness through
conservative backoff. All three strategies preserve aggregate throughput within 1\% of the
unmodified Aurora baseline. We identify an over-penalization failure mode in Strategy~A and Strategy~C in which stronger fairness
incentives produce worse outcomes due to interaction with the staggered evaluation
protocol.

The broader lesson is that RL-based congestion control fairness is a training-time problem
that cannot be fully addressed at deployment time. The three strategies studied here
represent complementary points in a design space bounded by reward engineering~(A),
information design~(B), and penalty calibration~(C), and together suggest that practical
improvements to RL-CC fairness are achievable without abandoning the RL framework.

% -----------------------------------------------------------------------
\balance
\bibliographystyle{abbrv}
\bibliography{references}

@inproceedings{jay2019aurora,
  author    = {Jay, Nathan and Rotman, Noga and Godfrey, Brighten and Schapira, Michael and Tamar, Aviv},
  title     = {A Deep Reinforcement Learning Perspective on Internet Congestion Control},
  booktitle = {Proceedings of the 36th International Conference on Machine Learning (ICML)},
  year      = {2019},
}

@inproceedings{liao2024astraea,
  author    = {Liao, Xudong and Tian, Han and Zeng, Chaoliang and Wan, Xinchen and Chen, Kai},
  title     = {Astraea: Towards Fair and Efficient Learning-based Congestion Control},
  booktitle = {Proceedings of the 19th European Conference on Computer Systems (EuroSys)},
  year      = {2024},
}

@article{brakmo1995tcpvegas,
  author  = {Brakmo, Lawrence S. and Peterson, Larry L.},
  title   = {{TCP Vegas}: End to End Congestion Avoidance on a Global {Internet}},
  journal = {IEEE Journal on Selected Areas in Communications},
  volume  = {13},
  number  = {8},
  pages   = {1465--1480},
  year    = {1995},
}

@misc{yen2024wififairness,
  author = {Yen, H.-C. and others},
  title  = {On the Fairness of {Internet} Congestion Control over {WiFi} with Deep Reinforcement Learning},
  note   = {arXiv preprint},
  year   = {2024},
}

@article{cardwell2016bbr,
  author  = {Cardwell, Neal and Cheng, Yuchung and Gunn, C. Stephen and Yeganeh, Soheil Hassas and Jacobson, Van},
  title   = {{BBR}: Congestion-Based Congestion Control},
  journal = {ACM Queue},
  volume  = {14},
  number  = {5},
  year    = {2016},
}

@techreport{jain1984fairness,
  author      = {Jain, Rajendra and Chiu, Dah-Ming and Hawe, William},
  title       = {A Quantitative Measure of Fairness and Discrimination for Resource Allocation in Shared Computer Systems},
  institution = {Digital Equipment Corporation},
  number      = {DEC-TR-301},
  year        = {1984},
}

@article{ha2008cubic,
  author  = {Ha, Sangtae and Rhee, Injong and Xu, Lisong},
  title   = {{CUBIC}: A New {TCP}-Friendly High-Speed {TCP} Variant},
  journal = {ACM SIGOPS Operating Systems Review},
  volume  = {42},
  number  = {5},
  pages   = {64--74},
  year    = {2008},
}

@article{liu2026efficient,
  title={Efficient Imputation for Patch-based Missing Single-cell Data via Cluster-regularized Optimal Transport},
  author={Liu, Yuyu and Yang, Jiannan and Yu, Ziyang and Pan, Weishen and Wang, Fei and Ma, Tengfei},
  journal={arXiv preprint arXiv:2601.14653},
  year={2026}
}

@inproceedings{liu2026hypehr,
  title={HypEHR: Hyperbolic Modeling of Electronic Health Records for Efficient Question Answering},
  author={Liu, Yuyu and Patil, Sarang Rajendra and Xu, Mengjia and Ma, Tengfei},
  booktitle={Findings of the Association for Computational Linguistics: ACL 2026},
  pages={10849--10862},
  year={2026}
}

@article{liu2026hyperguide,
  title={HyperGuide: Hyperbolic Guidance for Efficient Multi-Step Reasoning in Large Language Models},
  author={Liu, Yuyu and Xu, Haotian and He, Yanan and Patil, Sarang Rajendra and Xu, Mengjia and Ma, Tengfei},
  journal={arXiv preprint arXiv:2605.24140},
  year={2026}
}

\end{document}